\documentclass[conference]{IEEEtran}
\IEEEoverridecommandlockouts
\usepackage{cite}
\usepackage{amsmath,amssymb,amsfonts}
\usepackage{algorithmic}
\usepackage{graphicx}
\usepackage{textcomp}
\usepackage{xcolor}
\def\BibTeX{{\rm B\kern-.05em{\sc i\kern-.025em b}\kern-.08em
    T\kern-.1667em\lower.7ex\hbox{E}\kern-.125emX}}
\begin{document}

\title{Proxemics and Permeability of the Pedestrian Group}

\author{
    \IEEEauthorblockN{Saleh Albeaik, Faisal Alsallum, and Mohamad Alrished}
    \IEEEauthorblockA{\textit{Center for Complex Engineering Systems} \\
    \textit{King Abdulaziz City for Science and Technology}\\
    Riyadh, Saudi Arabia}
}

\maketitle

\begin{abstract}
    People tend to walk in groups, and interactions with those groups have a significant impact on crowd behavior and pedestrian traffic dynamics. Social norms can be seen as unwritten rules regulating people interactions in social settings. This article studies people interactions with groups and the emergence of group proxemics. A game theoretic model is outlined to analyze the primary elements to posed questions. Group zones, zone occupancy counts and people clearance from the group are studied using naturalistic data. Analysis indicate potential presence of three different zones in addition to the public zone. People tend to remain in the public zone and only progressively get closer to groups, and those closer approaches happen in a low frequency and for brief amount of time. 
\end{abstract}


\section{Introduction}
    
    The pedestrian group is an essential social construct in the study and analysis of crowd behavior \cite{nicolas2023social}. In a crowd, people tend to walk in groups who share social ties and exhibit behavioral patterns such as shared direction, pace, as well as emergent structures such as formations organizing the relative physical positions they assume \cite{nicolas2023social, bandini2014towards}. Those behavioral patterns have significant influence on crowd efficiency and safety, which is of significant concern to several research practice areas such as traffic engineering \cite{chen2018social} and crowd evacuation \cite{von2017empirical, zhao2017analysis}.

    Interactions take many forms in such crowds; from active conversations to passive observation of others. The type and nature of those interactions vary by context, type of crowd, among other factors. When a crowd is physically present, physical interactions and coordinated movement play a significant role. Those physical interactions are bound by a range of governing factors and rules from rational attempts to avoid physical collision \cite{helbing1995social}, emotional factors such as affection and social ties bringing people closer together \cite{Moaiteq2019EmotionalCD}, all the way to social norms regulating unwritten codes of conducts in interacting with other people \cite{Xin2025SocialNA}. 

    Social norms are unwritten codes of conduct governing people's behavior, which tend to be implicitly shared, communicated and transferred \cite{Chung2016SocialNA, Legros2019MappingTS}. Those rules evolved in ways people can understand and navigate. However, in addition to the scientific interest in understanding and documenting human behavior, as we attempt to model and simulate human behavior in software, or build machines to interact with people in acceptable ways, formal studies to dissect and articulate such behavioral expectations is unavoidable \cite{Lawrence2025TheRO}. 

    In a previous article \cite{albeaik2025groupagency}, we attempted to review and articulate a model for the group as an emergent agent. In this article, we attempt to extend that model to include social interaction aspect to it. Specifically, we study physical movement of people from crowds datasets and focus on how they interact with groups around them. We give particular attention to the distances they maintain from those groups. Moreover, we hypothesize that proxemics emerge around the group as a collective, which is an extension to the pair-wise individual to individual proxemics reported and studied in literature \cite{hall1966hidden}, and we report our findings from empirical data.

\section{Background}

    The study of pedestrian group proxemics involves central concepts such as the definition and structure of the group itself, structure of interactions between people, drivers to human behavior, as well as social norms as shared mutual expectations and regulators to social behavior. This section attempts to give a brief overview for these concepts as a background before delving into the core of the article. 

    \subsection{The pedestrian group}
        The literature reported several definitions for the pedestrian group \cite{moussaid2010walking, zanlungo2014potential, stangor2015social}. Here, we assume that if a set of pedestrian walk like a group, they are a group. In \cite{albeaik2025groupagency}, the group as a collective emergent agent has been discussed. It has been argued that such an agent have capacity for intelligent collective action, and could be defined by a composite state space describing the agent itself and influencing its behavior. The article in \cite{albeaik2025groupagency} focused on group agency state and its connection to behavior of pedestrian who are members of the group. In this article, we argue that other pedestrian who are outsiders to this group acknowledge this agency and behave according to shared expectations that emerge from and within such complex interactions. 

        The article further explored a structure to this group as an agent and its defining state spaces. Of particular relevance to this article, we highlight group-agency state variable as well as group collective intentionality and capacity to make decisions.

    \subsection{Social norms as drivers to behavior and regulators for social interactions}
        Social norms are unwritten rules regulating human behavior in society and social settings \cite{Chung2016SocialNA, Legros2019MappingTS}. When people walk around other people, social interactions are bound to happen. People generally cooperate to avoid collision and exchange cues to signal goodwill. Social norms emerge as systems of shared expectations regulating such interactions and prescribing what is acceptable or expected from otherwise \cite{Rafe2024ExploringTC}; for instance, who should go left and who should go right when pedestrian paths cross. The consequences to violation of social norms can range from awkward interactions, to mishaps such as a collision between two pedestrian reacting in incompatible ways, all the way to long lasting social judgment and social sanctions.

        In this article, we pay particular attention to pedestrian proxemics \cite{hall1966hidden}, which is concerned about people's behavior and attitude about space around them. We present a brief introduction about it here, and then we attempt to extend the concept to include the case of pedestrian groups and study empirical data to support the hypothesis. 

        \subsubsection{Introduction to proxemics}
            
            Edward Hall coined the term \emph{proxemics} and defined it as ``the interrelated observations and theories of humans' use of space as a specialized elaboration of culture". The theory describes four physical zones (or territories) defined by growing distances around each person, as can be seen in Figure~\ref{fig:individual-and-group-proxemics-representation} (top left). With those hidden unwritten rules for spaces around a person, only socially close people are welcome within the intimate zone, while generally close people can enter the personal zone, followed by generally familiar people who are allowed in the social space. Otherwise, general public are only permitted within the public space. 

            The concept of group proxemics has been investigated in literature with most attention being paid to detailing the classical proxemics theory. For instance, the authors of \cite{moussaid2010walking} explored proxemics and their impact on shape of group formation, the authors of \cite{bandini2014towards} explored proxemics dispersion as average distances people maintain between each other as they walk in group, and in \cite{gorrini2014group, manenti2010towards} focus was given to studying the effect of proxemics on crowd and its traffic flow dynamics. Within robot-human interactions, the authors of \cite{Samarakoon2022ARO, Neef2023WhatIA, Xu2023UnderstandingDH} studied appropriate (safety, comfort, acceptability, etc) distance robots are expected to maintain from people (as individuals).

            It could be noticed that proxemics are structured around interactions between individuals and details are specified in terms of social relationships between them. In what follows, we explore the situation when an individual is part of a bigger and more complex social entity such as a group. We study the nature of such interactions and and explore associated proxemics. 
    
\section{Structure of social norms in group interactions and the group proxemics hypothesis}
\label{sec:proxemics-hypothesis}

    We consider the pedestrian group as it interacts with other pedestrian in a crowd setting. We start by detailing the structure of interactions in those settings, the emergence of norms and shared expectations, and then detail the group-proxemics hypothesis. Towards the end of the article, we provide empirical data to support and detail the arguments of this hypothesis. 

    \subsection{Structure of interactions between the pedestrian group and outsiders}

        Social behavior can be seen in the patterns of walking and decisions people make as they navigate crowd situations. Interactions happen even when little attention is being paid to it by those people in crowd. This could be in how they share the physical space, and in the cooperation or lack thereof as two people pass each other to avoid collision. 

\begin{figure*}[!htbp]
            \centering
            \includegraphics[width=\linewidth]{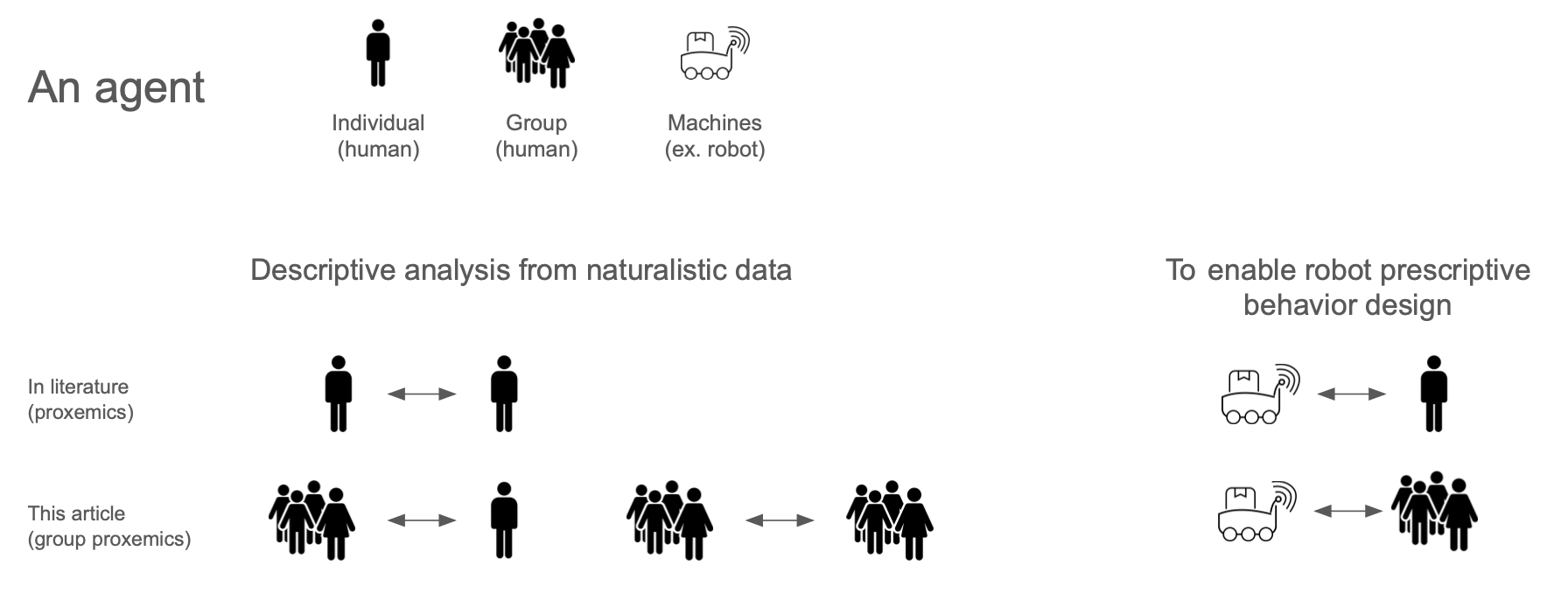}
            \caption{People as individuals as well as groups could be seen as agents. With the emergence of intelligent machines, those could be seen as agents as well. Multi-agent interactions between any of such agents influence their behavior and would be essential in the design of intelligent behavior in machines.}
            \label{fig:pairs-of-interaction}
        \end{figure*}

        Here, we focus on interactions and social norms that emerge around groups. We assume two sides to an interaction: an ego-group and an outsider as presented in Figure~\ref{fig:pairs-of-interaction}. The outsider could be an individual, a group, or a machine (such as a vehicle or a robot). As will be shown later in this article, groups interact differently with other groups compared to how individuals interact with groups. Furthermore, behavior in interactions with machines is expected to carry its own differences as well \cite{Lawrence2025TheRO, Samarakoon2022ARO, Neef2023WhatIA, Xu2023UnderstandingDH}. Those differences would be of significant importance as more machines are allowed to integrate with pedestrian traffic.
    
    \subsection{The emergent shared mutual expectations in interactions between the group and others}
        The group as an agent is expected to maintain its formation to be acknowledged as such. We argue that group proxemics emerge as a social norm with the emergence of the group as an agent. Within this social norm, the group is expected to protect their proxemics, while outsiders to the group are expected to respect those proxemics zones. 

\begin{figure}[!htbp]
            \centering
            \includegraphics[width=0.8\linewidth]{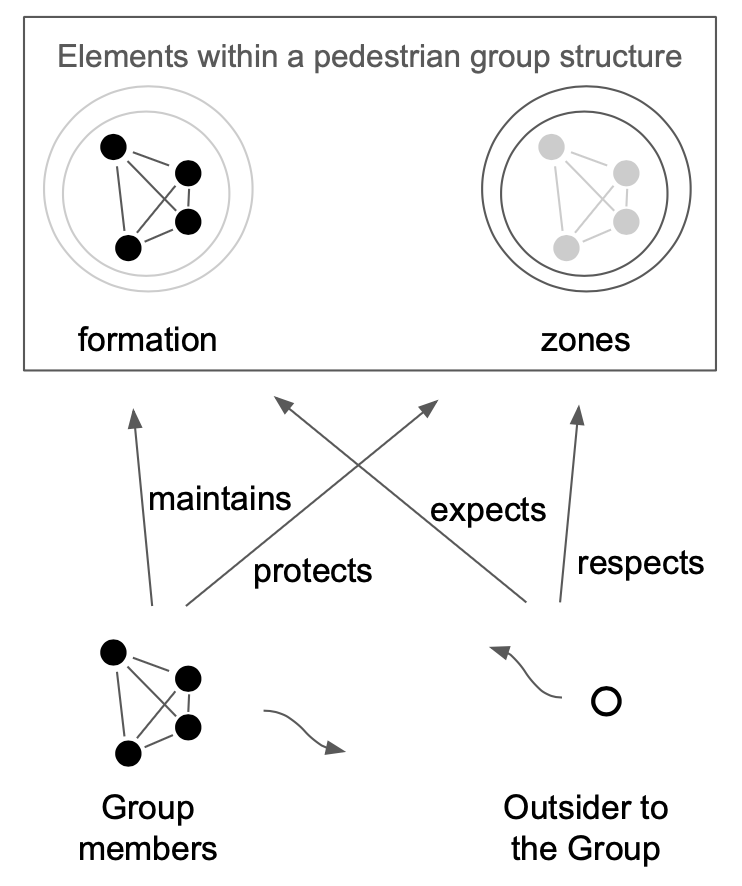}
            \caption{Structure of mutual expectations in the interaction between pedestrian and groups in crowd. When a group is formed, group proxemics emerge. The group is expected to maintain their formation and walk as to protect their zones, while outsiders to the group would avoid intrusions so long as the group maintain its part of this shared mutual expectations.}
            \label{fig:group-structure-of-social-norm}
        \end{figure}

        Here, we highlight two elements to this shared mutual expectations: expectations on formation, and expectations on zones as illustrated in Figure~\ref{fig:group-structure-of-social-norm}.  
    
        We hypothesize that members of a group and outsiders to the group have mutual expectations and behave accordingly. Group members will maintain a formation acting as a group in a detectable way, and they will collectively navigate (walk) and behave as to protect their own zones (while respecting the zones of outsiders to the group as well). On the other hand, outsiders to the group expect that the group is maintaining its formation, and in turn, the outsider will act in respectful ways to avoid violating group zones. 

    \subsection{Group Proxemics}
\begin{figure*}[!htbp]
            \centering
            \includegraphics[width=\linewidth]{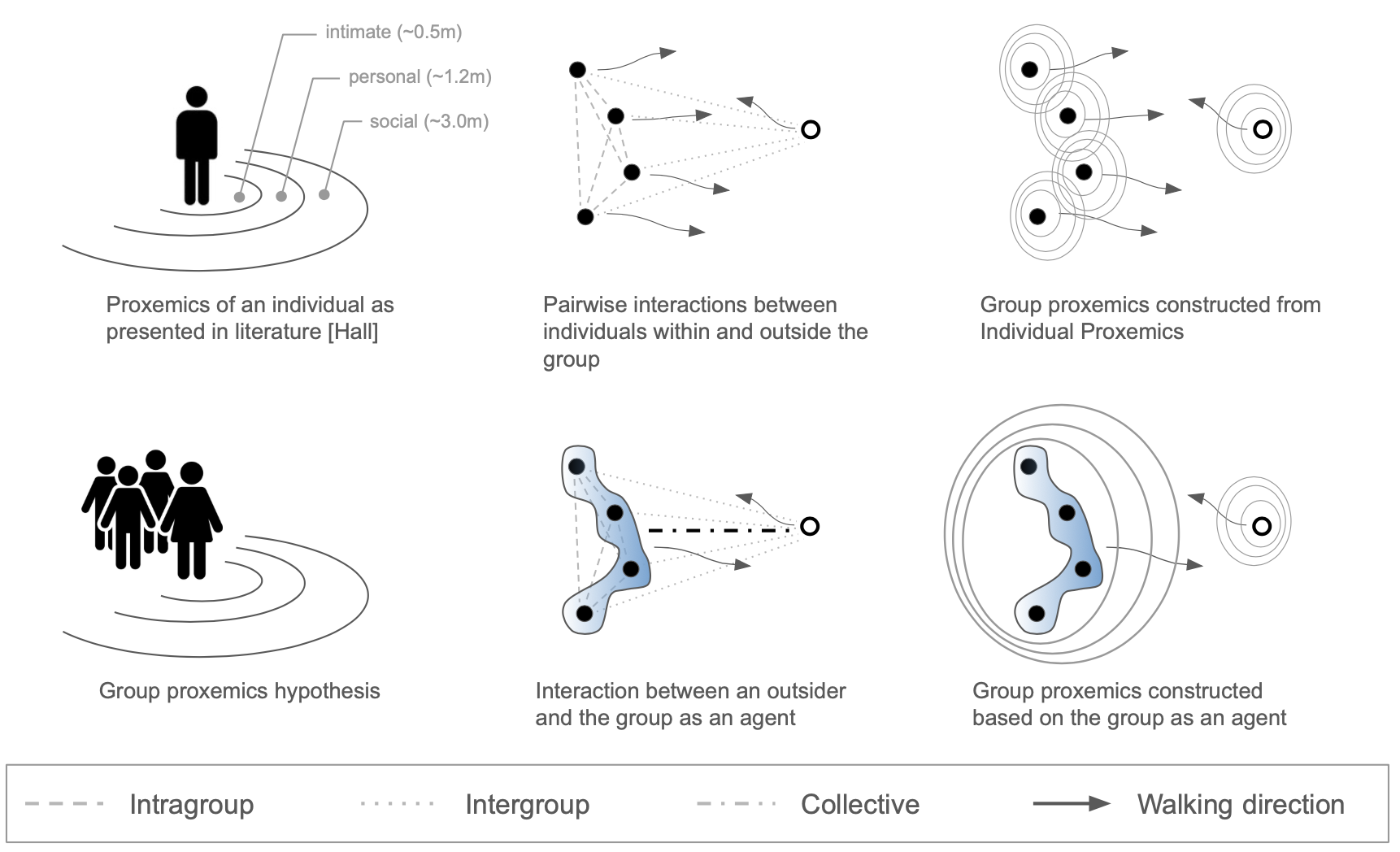}
            \caption{Hall \cite{hall1966hidden} coined the term ``proxemics'' in his studies of interactions between individuals. Group proxemics could be seen as an extension to account for proxemics in the interaction between outsiders with a pedestrian group as a collective.}
            \label{fig:individual-and-group-proxemics-representation}
        \end{figure*}

        Proxemics consider spaces around people and expectations around distances allowed or maintain between themselves and others. In this article, we highlight the distinction between group-to-outsiders proxemics as an extension to classical pair-wise proxemics. We also make distinction between proxemics internal to the same group (group member to group member of the same group, or intragroup proxemics), proxemics between a group member and an outsider to the group (group member to an outsider to the group, or intergroup proxemics), and more importantly for this article, group proxemics (between a group as a collective and an outsider to it). This is illustrated in Figure~\ref{fig:individual-and-group-proxemics-representation}.  

        \subsubsection{Group Proxemics derived for the group as an agent}
            In this article, we extend this concept of individual proxemics and hypothesize that upon the emergence of the group as an agent, group members and outsiders to the group start to observe an extension to this concept of proxemics as appropriate for the group as a collective. 
    
            Namely, as illustrated in Figure~\ref{fig:individual-and-group-proxemics-representation}, a set of zones emerge around the group as a collective. Similar to zones around individuals, zones around groups grow in size and are reserved progressively based on social relationships. However, as outlined in the following sections, those zones are more complex than zones of the individual. In addition to complexities that hold for both the group and individuals such as situational conditions and structure of the environment, group zones are maintained and observed by the coordinated action of independent decision makers.

    \subsection{Group Permeability}
        The proxemics zones defined in the previous subsection naturally invite two questions: as a zone defined around an emergent agent, when would this zone itself emerge and be acknowledged by others? and under what conditions, if at all, are people allowed to enter those zones? 

        To answer these questions, we propose to use the concept of \emph{group proxemics permeability} and define it as the extent to which social norm allows for outsiders to approach and enter into group zones (also referred to as intrusion and intrusion avoidance). Such term is important for the study of group proxemics as to help us differentiate between socially acceptable intrusions (or permeable situations) from empirically observed intrusions that otherwise are done in violation of social norms. 

        Note here the complexity emergence introduces to proxemics. In classical proxemics, zones surround a person, and they are always there. Thus, classical proxemics focuses on defining qualifications permitting another person to approach a zone. Additionally, upon impracticality or infeasibility (others lack of choice but to intrude) such as in situations of highly dense crowds, classical proxemics posits that such expectations are relaxed. However, permeability being discussed here extends beyond impracticality or physical infeasibility. As will be discussed later, it defines situations where intrusion avoidance is possible, but social expectation is relaxed. This could for example happen when a group moves in a way that violates the expectation on their formation or interacts with another group.  
        
\begin{figure*}[!htbp]
            \centering
            \includegraphics[width=\linewidth]{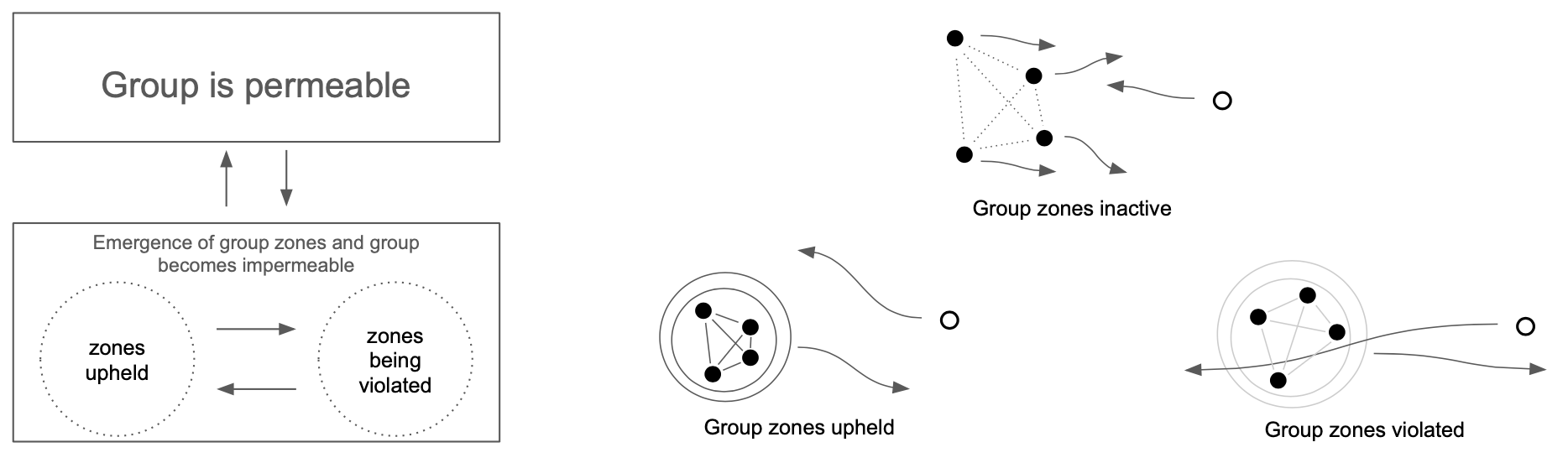}
            \caption{Group permeability indicates active zones with associated intrusion avoidance social norms from more flexible inactive zones.}
            \label{fig:group-permeability}
        \end{figure*}

        We thus highlight two notions: \emph{group permeability} and \emph{social norm violation} (or zone violation) as shown in Figure~\ref{fig:group-permeability}. We hypothesize that, when a group is permeable, people are allowed inside group zones as they are deemed inactive. Otherwise, people in general avoid such intrusions (being found within group zones) as they are considered a violation to social norm. 

        In interactions, there are at least two sides to a decision. While we have been paying most attention to behaviors and decisions of outsiders to the group, here, we highlight that a violation could happen when either of the two sides violate their part of this mutual expectation. For instance, group members could make an active decision to forego the expectation on who is allowed inside their zones. This happens for example when they actively decide to pass another slower pedestrian moving in the same direction ahead of them. This could be contrasted to an outsider decision to violate these expectations and actively decides to walk into group zones.

\section{Multi-agent Interaction Modeling}
    Assume a set of $N$ pedestrian walking within a 2-dimensional physical space for an observation duration of $T$ (with time variable $0 < t < T$). Each pedestrian is labeled $P_i$ for $i \in 1 \dots N$. Assume a set of $M$ groups present in the scene labeled as $G_j$ for $j \in 1 \dots M$. Each pedestrian $P_i$ is assumed to belong to one and only one group $G_j$. The physical position of each of the pedestrian at time $t$ is defined using the $2D$ vector $X_{P_i}(t) = \begin{bmatrix} x_{P_i}(t) & y_{P_i}(t) \end{bmatrix}$.
    
    We define group size (or group cardinality) function $\text{size}(G_j)$ as the number of members belonging to a group, with individuals assigned to groups of size 1. We also define group membership function $\text{belongs}(P_i)$ to extract the group this pedestrian belongs to, and conversely, we define a function to extract the list of all members to a group $\text{members}(G_j)$.

    Each dataset analyzed in this article is one continuous video recording as discussed later in this article. We dissect each dataset into group observations $\text{Obs}_{k}$ for $k \in 1 \dots K$, where $K$ is number of non-trivial groups in the dataset (of two members or larger). Each observation focuses on one ego-group from all groups observed. For each observation $\text{Obs}_{k}$, the ego group is defined as $G_1$ or $G_\text{ego}$. We define Agency State variable for the ego group $\text{Agency}_{G_\text{ego}}$, which takes a value from the set $\{ \text{Void}, \text{Transient}, \text{Active} \}$. 

    An interaction is defined by a pair of agents $A_1 \longleftrightarrow A_2$ where $A_1$ and $A_2$ can be any two agents observed in the dataset (including groups as agents). 

    \subsection{Interaction modeling}
        Multi-agent interactions are often modeled as games. A game is often modeled with: an agent set (or players), strategy set (or decision policies), and payoffs. In Game Theory, the objective is often to study strategy sets and identify equilibrium. Within our setting however, we assume players already know their strategies and our data represent outcome from the equilibrium reached at the social level. In this article, we can observe the outcomes from interactions but we do not know the strategies used explicitly nor the structure of payoff; analysis of such problems fall under Inverse Game Theory approaches.

        The problem studied in this article can be considered as a hierarchical multi-scale game \cite{Jin2021MultiScaleGR} representing intragroup interactions (players defined by the set $\{ P_h | h \in \text{members}(G_\text{ego}) \}$) and intergroup interactions (interactions between players from the set $\{ P_h | h \in \text{members}(G_\text{ego}) \}$ and players from the set $\{ P_h | h \notin \text{members}(G_\text{ego}) \}$) as shown in Figure~\ref{fig:individual-and-group-proxemics-representation}. We assume that, under appropriate conditions (when group agency state is active), the intergroup interactions can be abstracted and only effective collective behavior be considered. We thus consider the interacting pair $A_1 \longleftrightarrow A_2$; with $A_1$ representing an ego-group and $A_2$ representing an outsider to the group as players.

        We assume that those players engage in complex interactions and decisions. This article focuses primarily on studying the distance players maintains from each other and zone regulation. The distance problem can be formulated as a differential game while the zone regulation can be studied as a discrete decision and outcome game. 

        We assume potentially asymmetric decision spaces for each of the players to account for differences in agent complexity. That is, for the ego-group, we assume their decision space include (1) walking direction, in addition to (2) group agency and formation state decisions. On the other hand, when the outsider is an individual, their decision space is assumed restricted to (1) walking direction. 

        Outcomes are explored as the union of the physical positions of each of the players within the 2D space. Within the zone regulation problem, we discretize outcome to the set $\{ \text{No Intrusion}, \text{Strict Circle Intrusion}, \text{Hull Intrusion} \}$. 

        Within crowd behavior settings, we point out to external factors influencing behavior or outcomes, or restricting decision spaces. This includes physical constraints such as doorways or narrow passageways. 

        With this, we assume that payoffs represent social attitude towards different outcomes. Specifically, non-negative payoffs for maintaining social norms or outcomes not regulated by social norms, and negative payoffs for violating social norms. We assume that players are generally cooperative and respect social norms. We thus assume that frequently observed behavior is socially acceptable or is a social norm, and that the converse is true. That is, suboptimal behavior is assumed to be an indication for potential situations regulated by social norms. 
        
    \subsection{Group zone modeling}
        Here, we model the zones occupied by the group and areas around them to approximate potential zones groups actually maintain around themselves. We specifically consider the convex hull as tightest region and then a circle as a wider region. We specify these zones as follows:
        
        \subsubsection{Group center of mass}
            We estimate group center of mass (centroid) at time $t$ as $G_{\text{ego}, \text{centroid}}(t) = mean(X_{G_{\text{ego}}}(t))$ where $X_{G_{\text{ego}}}(t)$ is the set defined by vector position of each member of the ego group.
    
        \subsubsection{Group radius}
            We define group radius as the distance between centroid to the person within the group setting farthest from it. Thus we define group radius $ G_{\text{ego}, \text{radius}}(t) = \text{max}\{ \text{dist}( G_{\text{ego}, \text{centroid}}(t), X_{P_h} ) | P_h \in \text{members}(G_\text{ego}) \}$.

        \subsubsection{Group zones}
            We approximate group zones by two primary expanding zones; namely, the convex hull followed by a circle. The convex hull is defined around all members of the group as $\text{conv}({G_\text{ego}}) = \text{ConvHull}(\{ X_{P_h} | h \in \text{members}(G_\text{ego}) \})$. We further define the circle around a group as $\text{circle}(G_\text{ego})(t)$ with circle centroid $G_{\text{ego}, \text{centroid}}(t)$ and circle radius $G_{\text{ego}, \text{radius}}(t)$.

    \subsection{Interaction outcome measurement models}
        In this article, we use our data to estimate the following measurements: zone occupancy counts, and outsider distance distribution. 

        \subsubsection{Zone occupancy counts}
            We define zone occupancy counts as follows: for each zone around an ego group, zone occupancy count is the number of group observations $\text{Obs}_k$ within the whole dataset where this specific zone was found to be occupied by an outsider to the group (for some time during the observation; i.e., including brief occurrences). Thus, zone occupancy count could be estimated as $\sum_{k}{\text{Occupied}(\text{Obs}_k, \text{zone})}$. The function $\text{Occupied}$ is a zone occupancy indicator per group observation and is defined as follows: for some zone (such as the hull or circle defined above), $\text{Occupied}(\text{Obs}_k, \text{zone})$ is an indicator function and is active (value of one) if and only if $X_i(t) \in \text{zone}$ for some outsider $i \notin \text{members}(G_\text{ego})$ and some time $t$.

            To help us understand these intrusions, we classify them based on the different factors influencing pedestrian behavior. This includes, as detailed earlier, group agency state, decision attribution (to an outsider or to a group member), and priority conflicts (such as group to group interactions). Specifically for the following sections, we use the following indicators:
            \begin{itemize}
                \item Ego-group is in transience when $\text{Agency}_{G_\text{ego}} = \text{Transient}$.

                \item Ego-group initiated intrusion when, as defined earlier, ego-group is found to have made the decision to loosen its formation and walk in a way that brings an outsider into ego-group zones.

                \item Group-Group intrusion when for the pair of interaction $A_{G_\text{ego}} \longleftrightarrow A_2$ participating in an intrusion in $\text{Obs}_k$, $A_2$ is found to be a group; i.e., $\text{size}(A_2) > 1$.
            \end{itemize}

        \subsubsection{Outsider distance distribution}
            Occupancy counts gives insight into frequency of intrusions as a proxy to understand social behavior. However, to understand the distances people tend to maintain and clear around a group, and amount of time they spend at each distance, we propose to use outsider distance distribution discussed here.
            
            To estimate this distribution, for each group observation $\text{Obs}_k$, and at each time step $t$, we identify the outsider to the group walking closest to group centroid. That is, for each pedestrian $P_i$ for $i \notin G_\text{ego}$, we estimate $d_{P_i}(t) = \text{dist}( X_{P_i}(t), G_{\text{ego}, \text{centroid}}(t))$. Then at each time instance, we identify the pedestrian $P_\text{closest}(t)= \text{argmin}_{i}(d_{P_i}(t))$ and the distance this pedestrian is maintaining from the group $d_{P_\text{closest}}(t)= \text{min}_{i}(d_{P_i}(t))$. The curve defined by $d_{P_\text{closest}}(t)$ summarizes the distance around a group that is cleared (any outsider is present at this point or farther) at any point in time. We expect this distance to be affected by radius of the group. We thus normalize this function by group radius as ${d_{P_\text{closest}}(t)} / {G_{\text{ego}, \text{radius}}(t)}$.

            We then estimate the cumulative distribution function for the random variable $\text{Clearance}$ such that $F(\text{dist}) = P( {d_{P_\text{closest}}(t)} / {G_{\text{ego}, \text{radius}}(t)} \geq \text{dist})$. This is equivalent to the probability that there is no outsider to the group within $\text{dist}$ from group centroid; i.e., all outsiders are at $\text{dist}$ distance from the group or farther. In our analysis, we also condition on group size $\text{size}(G_\text{ego})$ to highlight potential differences in behavior.

\section{Naturalistic study on social norms and group interactions in crowds}
\label{sec:naturalistic-study}
    To analyze the inverse game problem discussed above, we conduct a naturalistic study based on observed human behavior. This article can be seen as an extension to the construction of the group-agent presented in \cite{albeaik2025groupagency}. The reader would find similarities and overlaps in the general structure and datasets used, however, the primary experiment and focus of the articles diverge in that this article is focused on studying interactions and social norms as opposed to the primary construction of the group agent. Next, we present the structure of the scientific approach we followed in this article.  

    \subsection{The general observed setup}
        In this article, we focus on studying the walking behavior of pedestrian in crowd as they interact with pedestrian groups. We approach our work as a naturalistic behavior study where we study behavior from surveillance video recordings of such crowds undisturbed. Such observational data has been documented and published in literature by several groups such as \cite{pellegrini2009you, bandini2014towards}. Those recordings are often preprocessed to generate pedestrian movement trajectories in addition to other information such as pedestrian group identification and labeling. 

        The following subsection presents the data used for the study in this article. We start from pre-processed crowd datasets with group labels.

    \subsection{Datasets}
        
\begin{figure}[!htbp]
            \centering
            \includegraphics[width=\linewidth]{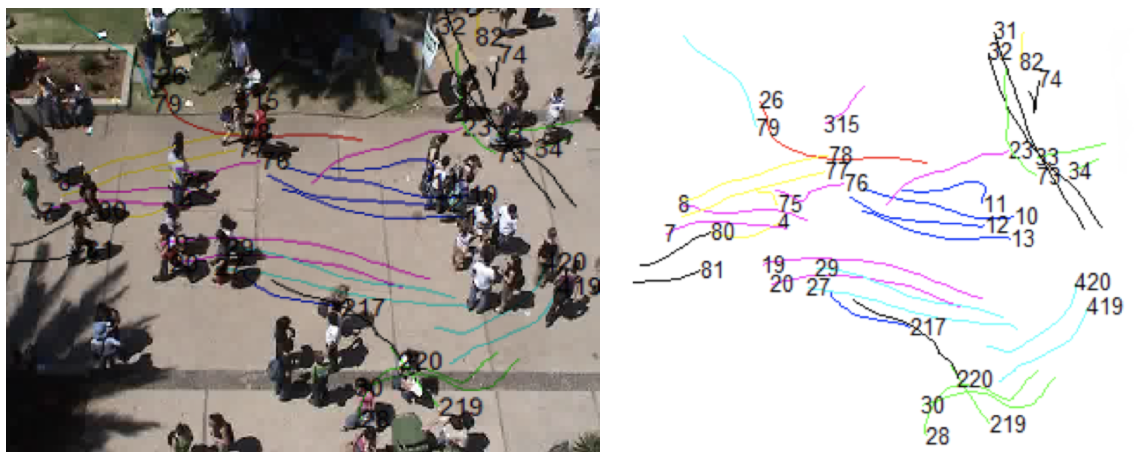}
            \caption{An image representing a general scene of observations with extracted pedestrian trajectories and group labeling (trajectory color). This is a sample from Student003 dataset \cite{lerner2007crowds}, with group annotation and illustration from \cite{solera2013structured}. This figure is borrowed from \cite{solera2013structured}.}
            \label{fig:dataset-example-image}
        \end{figure}

        For this study, we explore datasets that involve pedestrian groups within crowd traffic, and capture verity of interaction scenarios. Those tend to be crowd datasets with labeling of groups. Here, we focus on studying datasets collected from several crowd environments such as students at university, urban street, airport, and other public spaces. We focus on datasets being used heavily for pedestrian behavior studies, manually annotated, and captures different scenarios that vary along relevant dimensions that influence behavior such as crowd density, traffic types and directions, and demographics. Specifically, we use the following datasets:
        \begin{itemize}
            \item ETH-Univ and ETH-Hotel datasets \cite{pellegrini2009you}.
            \item GVEII dataset \cite{bandini2014towards}.
            \item Student003 dataset \cite{lerner2007crowds}.
            \item Collective Motion Dataset (CMD) \cite{zhou2013measuring}.
        \end{itemize}
        Trajectories and group annotations used in this work were conducted and published by \cite{amirian2020opentraj} and \cite{solera2013structured, solera2015socially}. A sample from such observation setup, along with a sample of the extracted trajectories and group labeling is shown in Figure~\ref{fig:dataset-example-image}.

        Although interactions between people in crowd and autonomous cars are starting to receive attention with new data being generated and published \cite{Yang2019TopviewTA, PaezGranados2022PedestrianRobotIO}. The data available is still limited or not fully labeled befitting to the requirements of the study in this article, we decide to focus on human-human interactions, and leave study of human-machine interactions as a future work.

    \subsection{Pre-processing, annotation and construction of the basic experimental event}
        The primary focus of this article is studying cases of interaction between groups and other people in the crowd scene. The primary observation in this article is a \emph{group observation}. We construct this observation by extracting the groups from each dataset, identifying the time frames where each group appeared in the dataset, and then extracting a copy of that part of the dataset to represent a group observation. A group observation as such is an annotated clip from a crowd scene where a specific group is observed from the time it enters the scene to the time it exits the scene. In \cite{albeaik2025groupagency}, we focus on the group itself and its walking behavior. Here on the other hand, we focus on its interaction with other pedestrian in the scene, and focus on behavior of those outsiders to this group.

\begin{figure}[!htbp]
            \centering
            \includegraphics[width=\linewidth]{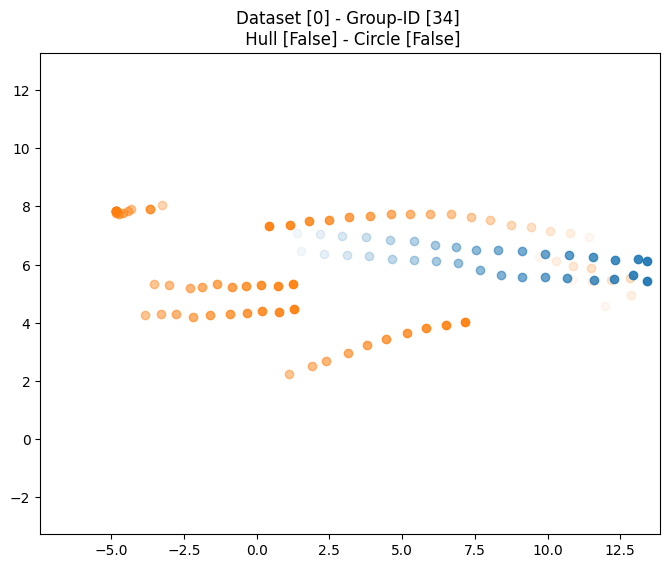}
            \caption{Sample trajectories extracted to represent a group observation along with other pedestrian in the scene.}
            \label{fig:sample-group-observation}
        \end{figure}

        From this dissection, we generated a video clip for each group observation as shown in Figure~\ref{fig:sample-group-observation} to review annotation of groups from original datasets to identify any mislabeling (groups labeled as a group but label appears to be incorrect) that might affect our quantitative results. We further annotated the data manual to label different aspects influencing interactions such as state of the ego-group (such as group being in a transient state), type of interaction (such as group interacting with another group, or conflict of priority), and ego-group decision to violate social norms during those interactions (such as decision to break group structure to pass another pedestrian).

        We focus on studying the distance people maintain from each other as they walk as well as regions of space they hold as their own. To analyze this behavior from empirical data, we assume a pedestrian group has a center of mass that could be computed by taking the geometric mean of the physical location of each of the members of the group. We also compute the convex hull defined around members of the group, as well as the circle defined by the center of mass of the group with radius equivalent to the member of the group farthest from the centroid. 
        
        We compute our counts and statistics by counting occurrences. The two primary events/measures we investigate are as follows: 
        \begin{itemize}
            \item \textbf{Frequency of intrusions:} We compute frequency of intrusions as frequency of outsider occupancy of each of the different hypothesized zones around the group as defined in Section~\ref{sec:proxemics-hypothesis}. Here, we assume the first and tightest zone is the convex hull, and the following zone is the circle around the group, then the remainder of open space around them. For each group observation, we evaluate if they experienced an intrusion into their convex hull at least once for the duration of observation, and if they experienced an intrusion into the circle around them. We use this to study how common intrusions are in general. 
            \item \textbf{The distance outsiders maintain from the group:} Here, we look into finer details. We compute the distance between group centroid and the outsider to the group walking closest to it. We take this measurement per time-step, and use it to construct a cumulative distribution function (CDF). This helps us understand how long occurrences of an intrusion last, and otherwise areas groups maintain clear most of the walking time. 
        \end{itemize}
        We use this data to provide empirical evidence and statistics to support and detail the group proxemics hypothesis. 

\section{Empirical results}
\label{sec:empirical-results}
    In this section, we study the hypothesis on the emergence of zones around the group. We present results of our analysis models of the datasets presented earlier. Here, we present quantitative analysis results and try to use it to support and refine the proxemics hypothesis presented in Section~\ref{sec:proxemics-hypothesis}. In attempting to infer social norms from observed behavior, we assume that people generally respect social norms. They rarely violate it, otherwise observed violations are assumed to be honest mistakes, or are explainable using details to the situation that are not apparent to the observer. 

    Therefore, in our quantitative study, we use this assumption as a basis to detect and infer the social norm from otherwise. Thus, what people do frequently enough is assumed an acceptable behavior. Otherwise, what people do most of the time (when an alternative action is possible and potentially even more optimal behavior but is rarely taken), we assume to be a social norm. On the other hand, what people rarely do (even when it is an optimal behavior), we assume to be a socially unacceptable behavior.

    \subsection{Decision structure hypothesis}
        Based on our preliminary expert evaluation, we assume agents employ a decision structure with the following determinants to intrusion:
        \begin{itemize}
            \item Zone activation state and permeability. We classify permeability into permeable, semi-permeable, progressively permeable/impermeable, and impermeable. We assume permeability is strongly correlated with group agency state. Zones are generally active when agency state is active. When group agency is active but formation is loose, zones are assumed semi-permeable in the sense that intrusions would be considered a social violation but further intrusions might still happen immediately after first case triggering the loose state. While the group is in a transient state, we assume that zones are progressively permeable/impermeable. 
            
            \item External environment factors and structural constraints such as entrances and narrow passageways. Formal analysis of this factor would require data and annotation not fully available to us. We thus decided to mention here but not emphasize further. 

            \item Multiple decision makers: the group and the outsider. A formal study of responsible decision maker from historical data would require formal causality analysis. To maintain focus of this article, we focus on key examples where a clear decision maker can be easily identified.

            \item Arbitration between agents of same status. For instance, when both sides of the interaction are a group, we assume that such pair of interaction would go through a different arbitration process than when the pair of interaction is clearly asymmetric such as an individual interacting with a group. 
            
        \end{itemize}

        This outline could be seen as a preliminary set of factors driving people within our setting. It could be used to eventually construct a more refined hypothesis on player strategies, which we leave as future work. In this article, we use these factors to explore the data and extract relevant measurements and statistics. 

    \subsection{Evidence on emergence of group zoning and progressive permeability}

        \begin{table*}[!htbp]
            \centering
            \begin{minipage}{.5\linewidth}
              \caption{Zone occupancy counts for ETH-Univ dataset.}
              \label{table:intrusion-counts-table-ETH-Univ}
              \centering
                \includegraphics[trim=6.8cm 16.2cm 6.8cm 1.8cm, clip=true, width=\textwidth]{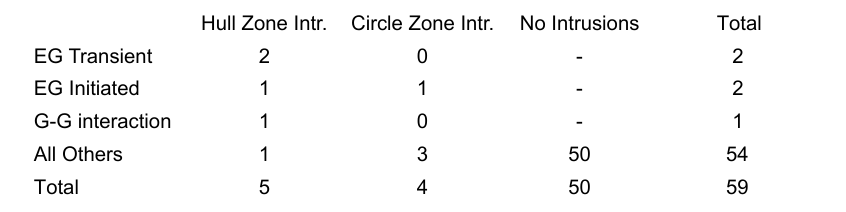}
            \end{minipage}%
            \begin{minipage}{.5\linewidth}
                \caption{Zone occupancy counts for GVEII dataset.}
                \label{table:intrusion-counts-table-GVEII}
                \centering
                \includegraphics[trim=6.8cm 16.2cm 6.8cm 1.8cm, clip=true, width=\textwidth]{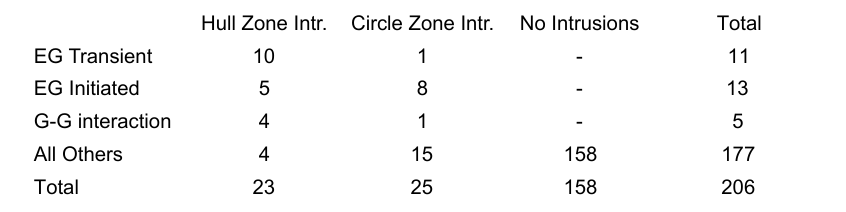}
            \end{minipage} 
        \end{table*}
        
        In this section, we study zone violation to better understand the extent to which zones are upheld, and responsibility to uphold those zones. We focus on two primary zones: namely, the hull as the tightest zone around a group, and then a circle as a wider area. 

        We split observed cases of intrusion into a four sets: intrusions while ego-group is in a transient state, intrusions initiated by the ego-group, and group-group intrusions. We assume that all other cases are initiated by an outsider to the group, and call them general cases. Here, we neglect other potential factors and causes to such violation such as visibility factors, crowd density, physical environment constraints, etc. 

        Furthermore, when we discuss outsider decision to intrude or not intrude, we count cases of ego-group in transient state and  ego-group initiated intrusions as no intrusion (basically, things are taken from perspective of the outsider to the group).

        As discussed earlier, in this section, we conduct event occurrence counting per group observation. We eliminate groups identified as mislabeling. Zone occupancy counts are presented in Table~\ref{table:intrusion-counts-table-ETH-Univ}~and~\ref{table:intrusion-counts-table-GVEII} for ETH-Univ and GVEII. ETH-Univ is sparse in the different types of intrusion rendering the counts unreliable for detailed statistics compared to GVEII. 

        Table~\ref{table:intrusion-counts-table-ETH-Univ}~and~\ref{table:intrusion-counts-table-GVEII} show that intrusions while ego-group is in a transient state represent 22\% and 23\% of all intrusion cases on ETH and GVEII dataset respectively. Group initiated intrusions represent 22\% and 31\% of all intrusions observed on the same two datasets respectively. Followed by 22\% and 10\% representing group-group intrusions. Otherwise, only 44\% and 40\% can be attributed to outsiders violating zone expectation not to intrude inside group zones. 
        
        Out of all observed groups, only 6.7\% (ETH) and 9\% (GVEII) groups experienced an outsider intruding into their zones. Those cases could further be classified into 1.9\% convex hull intrusions, and only 5\% and 7.3\% circle intrusions. When we eliminate ego-group attributed violations, outsiders thus respected group zones in 93\% and 91\% of the cases. 
        
        To summarize the findings reported in the table in terms of behavioral patterns:
        \begin{itemize}
            \item With near 100\% of the cases observed experienced hull intrusion while the ego-group was in a transient state, we say that a group is assumed permeable in this case and intrusions are expected. 
            \item Groups make decision to loosen expectations around their zones, and when they do so, hull intrusions and circle intrusions are both equally common. 
            \item When a group is in face to face interaction with another group, hull intrusions are frequently observed. We hypothesize that this is due to priority conflict with both agents of equal status as a group. 
            \item Outsiders generally respect group zones progressively; they avoid any intrusion, and if a situation makes it necessary, they intrude into the circle zone only, and they avoid intrusion into the hull at all costs and unless absolutely necessary. 
        \end{itemize}

    \subsection{Clearance attitude and emergence of the public zone}

        \begin{figure*}[!htbp]
            \centering
            \includegraphics[trim=0cm 1.45cm 0cm 0cm, clip=true, width=\linewidth]{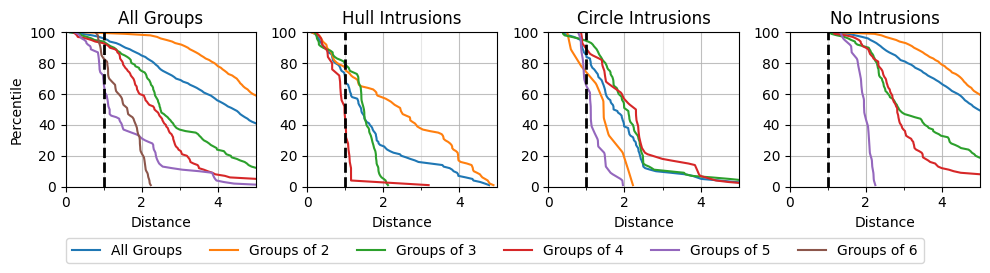}\vspace{0.25cm}
            \includegraphics[trim=0cm 1cm 0cm 0.7cm, clip=true, width=\linewidth]{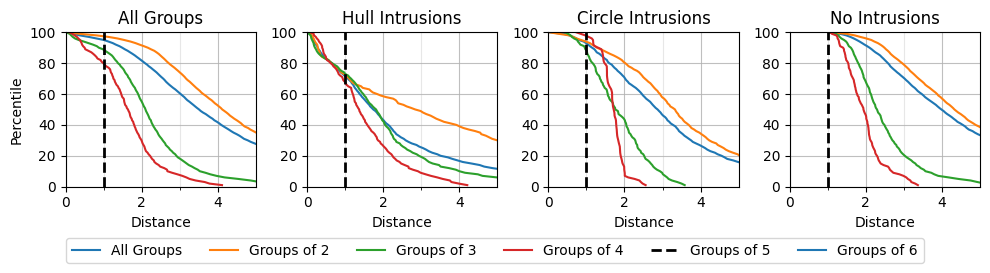}\vspace{0.25cm}
            \includegraphics[trim=0cm 0cm 0cm 6cm, clip=true, width=\linewidth]{figures/data-analysis-results/ped_approaches_cdf_dataset_0.png}
            \caption{Outsider clearance distance cumulative distribution functions (cdf). First row shows cdfs for ETH-Univ dataset and second row shows cdfs for GVEII dataset.}
            \label{fig:group-outsider-distances-distribution}
        \end{figure*}

        In the earlier section, we have noted the observation that violation of expectation not to cross between group members can be said to be a rare occurrence. We also noted that people avoid intruding into a circular zone around the group as well. 
        
        In this section, we study people clearance attitude in general, and we study the next emerging zone by focusing on cases of no violation of this intrusion avoidance expectation. By studying remaining data, we study probability distribution of closest approach to the group by outsiders of it. We notice a clear progressive behavior indicating a potential emergence of such zone. Namely, we observe that when outsiders to the group do not have to intrude into the group, they maintain at least $\text{group radius}+\delta$ (with $\delta$ being non negligible), regardless of how big group radius is and regardless of how many people are in the group as can be seen in Figure~\ref{fig:group-outsider-distances-distribution}. 

        In general, we notice an S-shaped curve. The area represented by $\text{radius}+\delta$ around the group is almost always cleared by other people in the crowd. We then notice a sharp change in probability distribution at after this bound. 

        To examine Figure~\ref{fig:group-outsider-distances-distribution} closely, we discuss each of the four plots. Each diagram is a CDF (cumulative distribution function) for measuring statistics of how close people tend to get to the group as they walked. The first plot on the left presents an overall summary for all groups observed. It is then followed by three plots presenting conditional CDFs (groups that experienced a hull intrusion, groups that experienced a circle intrusion, groups that did not experience any intrusion). Where the probability of each type of intrusion can be inferred from the earlier subsection.
        
        The horizontal axis in these plots is normalized distance (how far someone is to the group normalized by radius of the group); as such, distance of one is for someone exactly at group radius. The vertical axis represents CDF value. This chart answers the following question: how many frame (i.e., how long) were there someone inside the circle? and how many frames people where outside the circle? To present an example to read this plot; for groups of size five, about $80\%$ of the time, people where $1.8\times\text{group radius}$ far from the group or farther, people where $1.9\times\text{group radius}$ or farther for about $60\%$ of the time, and people where $2\times\text{group radius}$ or farther for about $45\%$ of the time.

        In general, we notice that people maintain a significant distance outside the circle of the group, or what can be said the public zone, regardless of how big the group is in cardinality or in radius. When people intrude into the circle of the group, they generally either intrude superficially (do not go deep into the circle) or spend as little time as possible inside. In cases of hull intrusion, people get deepest into the group. However, it can be noted that here too, people stay outside of the circle as well almost $70\sim80\%$ of the time.
    
\section{Conclusions}
    This article investigates pedestrian interactions with pedestrian groups. We hypothesize that shard expectations and social norms emerge around the group as a whole. We focus our attention on proxemics and hypothesize that people maintain significant distance from groups and observe different zone progressively. Multi-agent interactions are proposed to be modeled using game theory along with analytical models to estimate relevant metrics. Crowd trajectory datasets were used for empirical data analysis. Our analysis highlighted that outsiders intruded into groups in less than $10\%$ of the encounters, and when those intrusion happened, people spent less than $20\%$ of their walking time inside group zones.

\bibliographystyle{IEEEtran}
\bibliography{references}


\appendices

\section{Hand crafted predictive indicators to automate dataset annotation}
\label{sec:hand_crafted_indicators}
    The analysis in this article require additional labels not already available and published. With limited manual resources, we labeled two datasets (ETH-Univ and GVEII). To enable extension of the analysis beyond these two datasets, here, we propose simplified functions to automate the process of annotation based on conclusions presented in \cite{albeaik2025groupagency}. In the following appendix, we present results based on this automated labeling for all the datasets discussed in Section~\ref{sec:naturalistic-study}. 

    \subsection{Group transience}
        The group was observed to be in a transient state when group radius grows above $2m$. We thus implement the detection of transition between the active-agent state and the transient-agent state by: $\text{radius} = 2$. And thus, group is considered in a transient state when $\text{radius} > 2$, and group is in active-agent state when $\text{radius} \le 2$. We expect this filter to help us identify group observation instances that experience transience with high confidence (high confidence on true positives with high accuracy, high precision, but possibly low recall) to generate statistics for groups in transience. On the other hand, distribution of data is skewed with significantly more groups in active-agency state than in transience. Thus, we expect that the false negatives would be diluted by the larger number of true negatives to generate meaningful statistics for groups in active-agency state. 

        Estimated indicator accuracy on ETH-Univ dataset is $97\%$, and on the GVEII dataset is $96\%$. Note that we use those indicators and this accuracy evaluation as rough estimates rather than a formal study on development of predictive indicators and their accuracy. Moreover, we use these indicators as rough labels to help us explore datasets for which we do not have manual annotation.

    \subsection{Group initiated intrusion}
        We assume that a group may initiate intrusions in one of the following cases:
        \begin{itemize}
            \item \textbf{Ego-group passing a slower pedestrian moving in the same direction:} To test for ego-group passing a slower pedestrian, we test for direction similarity (difference between ego-group heading and outsider heading is less than $\pi /4 $) and group walking at faster speed than outsider.
            \item \textbf{Ego-group passing a stationary pedestrian approached from any direction:} This is tested by identifying an outsider intruding into the group and be found to be stationary (moving speed equal to zero). 
        \end{itemize}

        Estimated indicator accuracy on ETH-Univ dataset is $97\%$, and on the GVEII dataset is $93\%$. Note that we use those indicators and this accuracy evaluation as rough estimates rather than a formal study on development of predictive indicators and their accuracy. Moreover, we use these indicators as rough labels to help us explore datasets for which we do not have manual annotation.

    \subsection{Group-to-group type intrusion}
        This is tested by examining intrusions into a group, and testing if the intruder is identified as a pedestrian-group. 

        Estimated indicator accuracy on ETH-Univ dataset is $93\%$, and on the GVEII dataset is $90\%$. Note that we use those indicators and this accuracy evaluation as rough estimates rather than a formal study on development of predictive indicators and their accuracy. Moreover, we use these indicators as rough labels to help us explore datasets for which we do not have manual annotation.

\section{Replication of Analysis on large datasets based on automatic labeling}
    We use the the hand crafted indicators presented in Appendix~\ref{sec:hand_crafted_indicators} to automatically label datasets discussed in Section~\ref{sec:naturalistic-study}. We replicate the statistics we presented in Section~\ref{sec:empirical-results}. 
    
    Zone occupancy counts---reported as rate, count / total number of observations, to enable direct comparison between datasets---are presented in Table~\ref{table:intrusion-counts-all-datasets-automated-labeling-table}. For compactness, in this table, occupancy counts are classified into: permeable groups (intrusions are possible but are not considered a social norm violation), impermeable groups experiencing intrusions (social norm violation), and groups that experienced no intrusion. Social norm violating intrusions (here, defined as the union of hull and circle intrusion) are classified further into: ego-group initiated intrusion, group-group interactions, and potentially outsider initiated violation. 
    
\begin{figure*}[!htbp]
        \centering
        \includegraphics[width=0.45\linewidth]{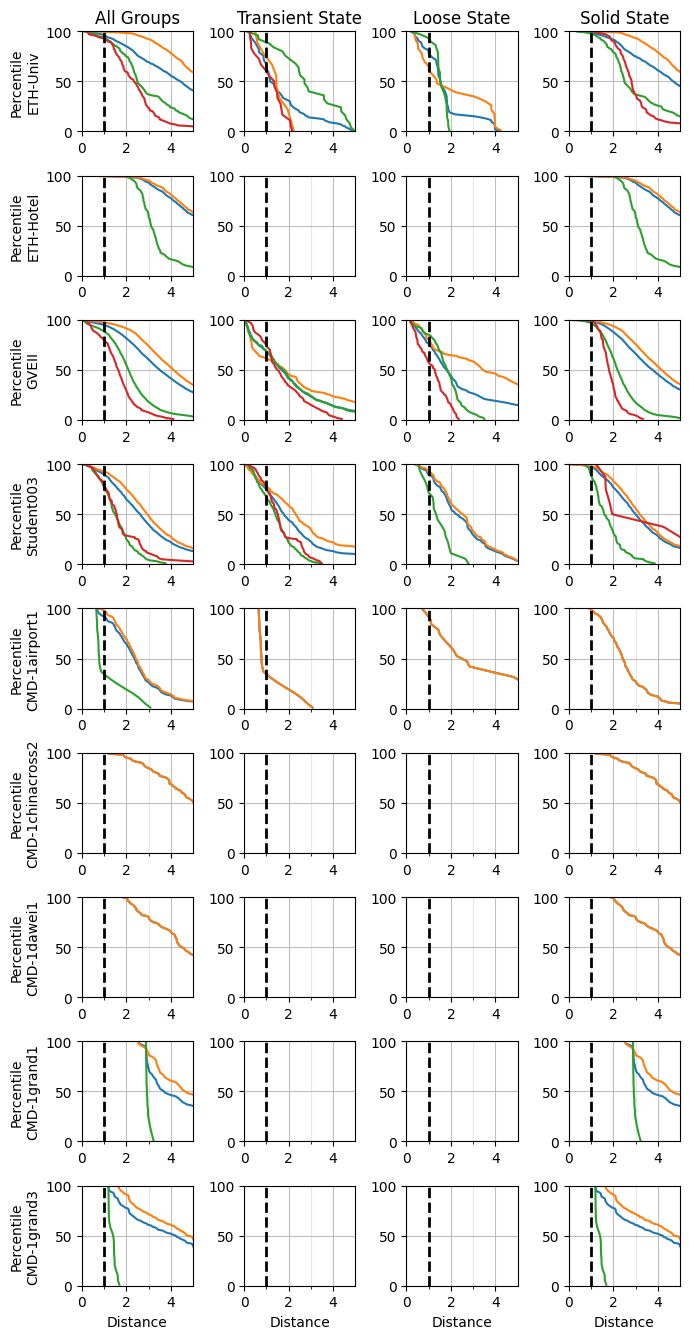}
        \includegraphics[width=0.45\linewidth]{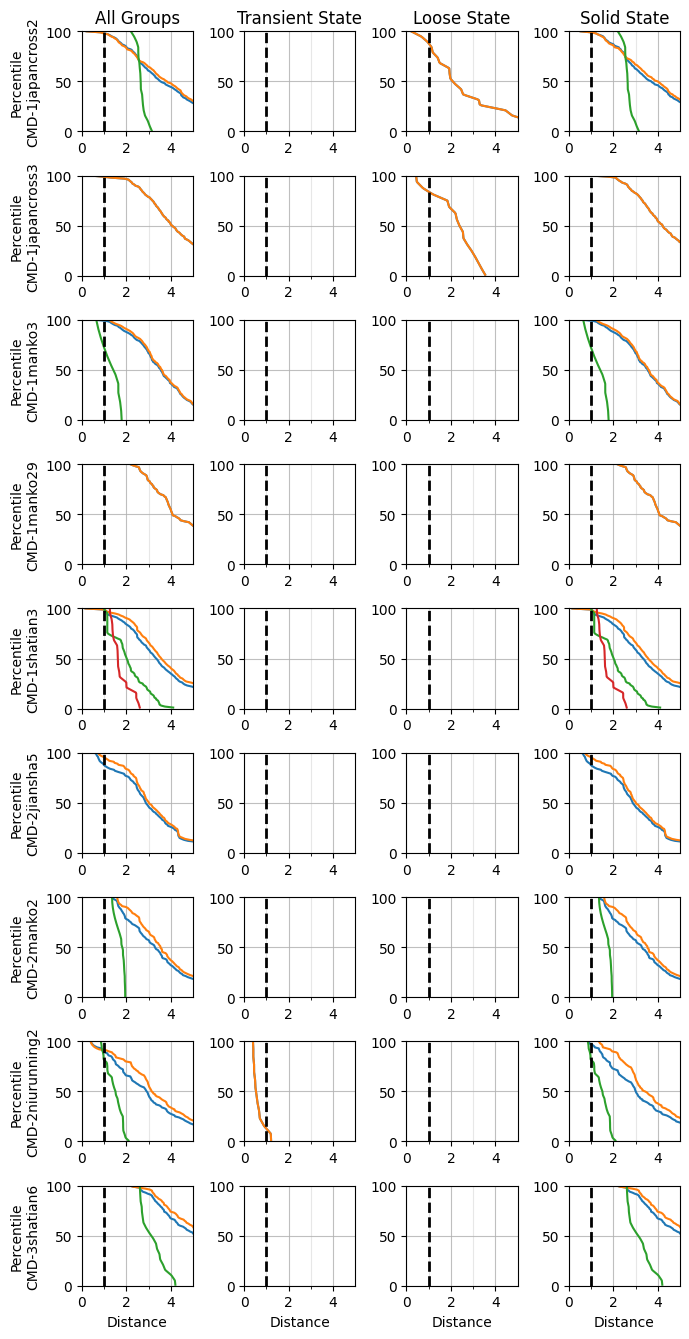}\vspace{0.25cm}
        \frame{\includegraphics[trim=1.7cm 0.3cm 9.2cm 6.1cm, clip=true, width=0.6\linewidth]{figures/data-analysis-results/ped_approaches_cdf_dataset_0.png}}
        \caption{Outsider clearance distance cumulative distribution functions (CDF). CDF is shown for each of the datasets discussed in Section~\ref{sec:naturalistic-study}. Data presented here is based on automated labeling as discussed in Appendix~\ref{sec:hand_crafted_indicators}.}
        \label{fig:cdfs_all_datasets_automated_labeling}
    \end{figure*}

    Cumulative distribution functions (CDFs) are presented in Figure~\ref{fig:cdfs_all_datasets_automated_labeling}. We leave these results to the reader to explore consistencies or lack thereof from detailed analysis in the main sections of this article. 

    \begin{table*}[!htbp]
        \centering
        \caption{Intrusion rates observed from each of the datasets discussed in Section~\ref{sec:naturalistic-study}. Group Observations are classified into cases that were deemed permeable or experienced no intrusion in addition to three different types of intrusion (ego-group initiated, group-group interaction, and those potentially an outsider violation to social norm). Data presented here is based on automated labeling as discussed in Appendix~\ref{sec:hand_crafted_indicators}.}
        \label{table:intrusion-counts-all-datasets-automated-labeling-table}
        \includegraphics[trim=1cm 2.5cm 1cm 2.5cm, clip=true, width=\textwidth]{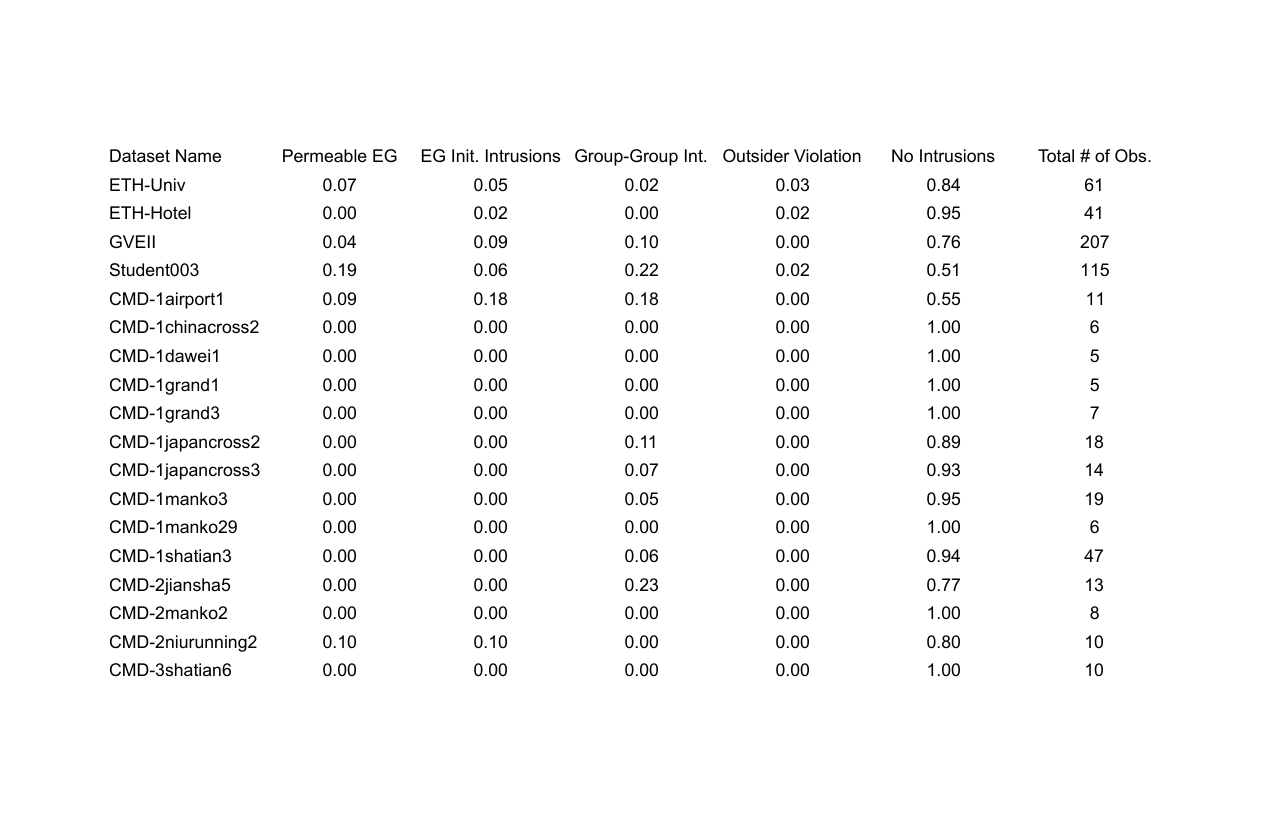}
    \end{table*}

\end{document}